\documentclass[twocolumn,showpacs,preprintnumbers,amsmath,amssymb,prl]{revtex4}
\usepackage{graphicx}
\usepackage{dcolumn}
\usepackage{bm}
\usepackage[tight]{subfigure}
\usepackage{amsmath}
\usepackage{verbatim}
\usepackage{soul,color}
\newcommand{\half}{\textstyle \frac{1}{2}}

\begin{document}
\title{
Pair-tunneling resonance
in the single-electron transport regime
}
\author{M. Leijnse$^{(1,3)}$}
\author{M. R. Wegewijs$^{(1,2,3)}$}
\author{M. H. Hettler$^{(4)}$}
\affiliation{
  (1) Institut f\"ur Theoretische Physik A,
      RWTH Aachen, 52056 Aachen,  Germany \\
  (2) Institut f\"ur Festk{\"o}rper-Forschung - Theorie 3,
      Forschungszentrum J{\"u}lich, 52425 J{\"u}lich,  Germany \\
  (3) JARA- Fundamentals of Future Information Technology\\
  (4) Institut f{\"u}r Nanotechnologie, 
      Forschungszentrum Karlsruhe, 76021 Karlsruhe, Germany \\
}
\begin{abstract}
  We predict a new electron pair-tunneling (PT) resonance
  in non-linear transport through quantum dots
  with positive charging energies exceeding the broadening due to thermal and quantum fluctuations.
  The PT resonance shows up 
  in the \emph{single-electron} transport (SET) regime
  as a peak in the derivative of the
  non-linear conductance, $d^2I/dV^2$,
  when the electrochemical potential of one electrode matches the average of
  two subsequent charge addition energies.
  For a single level quantum dot (Anderson model) we find the analytic peak shape
  and the dependence on temperature, magnetic field and junction asymmetry
  and compare with the inelastic cotunneling peak which is of the same order of magnitude.
  In experimental transport data the PT resonance 
  may be mistaken for a weak SET resonance
  judging only by the voltage dependence of its position.
  Our results provide essential clues to avoid such erroneous interpretation of transport spectroscopy
  data.
\end{abstract}
\maketitle
{\em Introduction.}
Electron tunneling spectroscopy 
has nowadays become a standard tool for investigating the in situ properties of 
single-electron transistors based on quantum dots in
semi-conductor hetero-structures, nanowires, carbon nanotubes and even single
molecules. 
The basic spectroscopy rules derive from the simple energy resonance conditions for
single-electron tunneling (SET) onto the {dot}.
As a result, the bias positions of differential conductance ($dI/dV$) resonances
depend linearly on the gate voltage due to capacitive
effects~\cite{Kouwenhoven97rev}.
In addition, inelastic cotunneling (COT) processes
can often  be distinguished as steps in $dI/dV$~\cite{Lambe68},
 allowing for spectroscopy of quantum dot excitations with increased energy 
resolution~\cite{DeFranceschi01}.
Here an electron is transferred from the high to the low biased electrode
through the dot, using the excess bias-energy to 
excite the dot by an energy $\Delta \epsilon$. 
Since these electron-hole charge transfer resonances involve only a virtual charging of the dot,
they appear at a bias threshold $V=\Delta \epsilon$, independent of the gate voltage.
These processes 
arise only in second order perturbation theory in the tunnel rate $\Gamma$.
\par
In this letter we show that, in the same order of $\Gamma$,
electron pair-tunneling (PT) gives rise to distinct, measurable
transport effects which, to our knowledge, have been overlooked so far.
In a single coherent process, a pair of electrons is extracted from (added to) one
electrode, {resulting in a real (dis)charging of the dot by two electrons, 
involving a positive charging energy much larger than broadening due to thermal and quantum charge fluctuations.}
The corresponding PT resonance appears deep inside the SET region where a finite 
current is flowing and its position has the same gate-dependence as SET resonances. 
It might thus be mistaken for a weak SET resonance belonging to some excited state and
one might thereby extract an erroneous level-structure from the spectroscopic data.
We show how the PT 
resonance can be distinguished from other types of first and second order processes and what additional 
information can be extracted from it.
Since spectroscopic data obtained from single-electron transistor devices are often 
non-trivial to interpret 
{we provide several independent criteria for its experimental identification, namely  
the resonance shape and the dependence on temperature, magnetic
field and junction asymmetry}. 
{The PT processes discussed here} completely lack any signature inside the Coulomb blockade region 
where SET is exponentially suppressed. {In this region a different type of pair-tunneling resonance
occurs which is much weaker than the one discussed here since it appears only in the third order 
in the tunnel rate $\Gamma$}~\cite{Sela08}. {Also the latter 
pair-tunneling process involves neither real nor virtually doubly occupied states (i.e. remains 
finite in the $U = \infty$ limit)}.
Second order pair-tunneling resonances were discussed {previously}
for the rare case of \emph{negative} Coulomb charging energy (i.e. effective attractive
electron-electron interaction)~\cite{Koch05c}, related to
pair-tunneling in superconducting grains~\cite{Hekking93},
where SET is suppressed by the superconducting gap.
{Also Kondo effect involving electron pairs have been discussed in this limit}~\cite{Cornaglia05b}.
{Here we consider the experimentally most frequently occurring case of strong positive charging energy.}
\par
{\em Model.}
The pair-tunneling resonance is already present in the generic model for a quantum dot, i.e. the non-equilibrium Anderson model.
Here the quantum dot, described by the Hamiltonian 
$H = \sum_{\sigma} \epsilon_{\sigma} {\hat n}_{\sigma} + U {\hat n}_{\uparrow} {\hat n}_{\downarrow}$,
consists of a single orbital {with energy $\epsilon_{\sigma}$ and occupation 
${\hat n}_{\sigma} = d^{\dagger}_{\sigma} d_{\sigma}$.
Here $\epsilon_{\downarrow} - \epsilon_{\uparrow} = h$ equals the Zeeman energy and $U$ is the 
finite positive charging energy.} {We let $N = \sum_{\sigma} {\hat n}_{\sigma}$ denote the 
electron number on the dot.}
The \emph{many-body} eigenstates are $|0\rangle$, $|\sigma \rangle$ and $|2\rangle$ with 
energies $E_0 = 0$, $E_{\sigma} = \epsilon_\sigma$ and $E_2 = \sum_\sigma \epsilon_\sigma + U$.
The dot is coupled by a tunneling Hamiltonian, 
$H_\text{T} = \sum_{r=L,R}\sum_{k,\sigma} T_{r} d_{\sigma}^\dagger c_{rk\sigma} + h.c.$, 
to macroscopically large reservoirs, described by 
$H_\text{R} = \sum_{r=L,R}\sum_{k,\sigma} \epsilon_{k}c_{rk\sigma}^\dagger c_{rk\sigma}$. 
The electrons in the reservoirs are assumed to be non-interacting, with operators $c_{rk\sigma}^\dagger,c_{rk\sigma}$
for state $k$ and spin $\sigma$ in electrode $r=L,R$.
The tunnel amplitudes can be expressed in the tunnel rates $\Gamma_r$ through $T_{r} = \sqrt{\Gamma_{r}/2\pi}$.
Throughout the paper we use natural units where $\hbar = k_\text{B} =|e| = 1$, where $-|e|$ is the electron charge.
\par
\begin{figure}[t!]
     \includegraphics[height=0.4\linewidth]{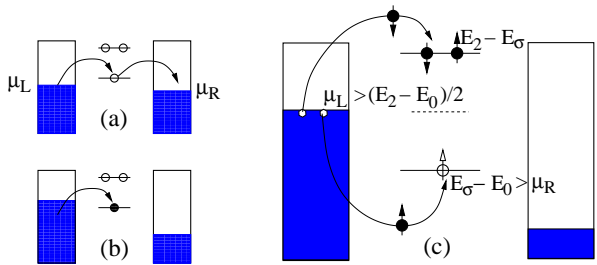}
    \caption{
    \label{fig:resonance}
     (Color online).
     Energy differences 
     between \emph{many-body} eigenstates (dot chemical potentials) of the Anderson model and 
     sketch of a COT process (a) and a SET process (b).
    (c): More detailed sketch of a PT process, giving rise to a resonance at $\Delta = U/2$ 
    above the SET onset. Filled (unfilled) circles indicate real (virtual) occupation.
}
\end{figure}

Let us first present the basic physics.
Fig.~\ref{fig:resonance} 
shows a schematic comparison of tunneling processes
for the spin-degenerate Anderson model when $E_{\sigma}$ is larger than the average
chemical potential $\mu$ of the reservoirs, such that the quantum dot is unoccupied at zero bias voltage, $V = 0$.
We take $\mu = 0$ and assume symmetric biasing such that $\mu_\text{L} = V/2$, $\mu_\text{R} = -V/2$. 
At low bias transport is dominated by cotunneling involving \emph{virtual} occupation
of state $|\sigma \rangle$. For larger bias, $|V/2| > E_\sigma - E_0$, electrons can sequentially tunnel
into and out of the dot, involving \emph{real} occupation of state $|\sigma \rangle$.
In this regime double occupation of the dot through two consecutive SET processes only becomes energetically 
allowed when additionally the energy difference $E_2 - E_\sigma$ is below the chemical potential of 
one lead, i.e. when $|V/2| > E_{\sigma} + U$.
Midway between these resonances the coherent
tunneling of a \emph{pair} of electrons from the same reservoir becomes possible, i.e. at the resonance condition: 
\begin{equation}
\label{eq:resonance_condition}
	 |V/2| > \half \left( \sum_{\sigma}E_{\sigma} + U \right).
\end{equation}
The corresponding process is shown in Fig.~\ref{fig:resonance}(c). 
One can think of this as one electron from just below the Fermi level of the left
reservoir tunneling onto state $| \sigma \rangle$, leaving an excess energy 
$\Delta = U/2$, which can be used to assist the second electron in reaching state
$| 2 \rangle$. Thus, the \emph{total} PT process is energy-conserving. 
However, since there are no internal degrees of freedom on the dot to 
store the energy $\Delta$, these two processes have to take place 
coherently in the short time set by the time-energy uncertainty relation.
\par
Experimentally, the dot energies can be
controlled linearly by the gate voltage $V_\text{g}$. 
This produces the characteristic  Coulomb diamond figures in differential
conductance color-maps as function of gate- and bias voltage~\cite{Kouwenhoven97rev}.
Since the resonance condition~(\ref{eq:resonance_condition}) for
PT depends on the \emph{average} of the two dot energies $E_\sigma$ the PT lines have the 
same slope as SET resonances.
To distinguish the PT and SET resonances a consistent calculation of their transport signature is
thus imperative.
\par
{\em Transport calculation.}
The pair-tunneling resonance is visible in the transport spectrum when the charging energy is larger than 
both the tunnel rate $\Gamma$ and temperature $T$, and originates from coherent two-electron processes 
which become important with increasing {$\Gamma$}.
We can address the interesting regime $U > T > \Gamma$  using the real-time transport 
theory~\cite{Koenig96prl}, {which allows systematic treatment of processes beyond 
lowest order in $\Gamma$, while accounting for the competition of single- and two-electron 
processes, essential at the PT resonance}.
The central task is to calculate the stationary reduced density matrix of
the dot in the basis of many-body eigenstates, which is done using a 
kinetic (master) equation. From this density matrix the current flowing out of 
reservoir $r$, $I_r$, can be calculated.
The required transport rates are calculated perturbatively up to second order in $\Gamma$ and
are given by the analytic expressions which we derived for a general model in Ref.~\cite{Leijnse08a}.
Importantly, \emph{all} coherent one- and two-electron processes are included and the Coulomb 
interaction $U$ is treated \emph{non-perturbatively}.
\par
\begin{figure}[t!]
      \includegraphics[height=0.85\linewidth]{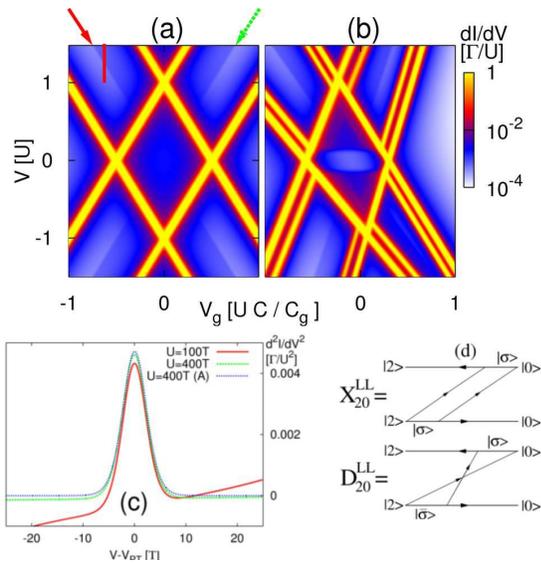}
  \caption{
    \label{fig:anderson}
    (Color online).
    (a): $dI/dV$ vs $V,V_g$ plotted on logarithmic color scale for $\Gamma_\text{L} = \Gamma_\text{R} = \Gamma$,
    $U = 100T =  2000 \Gamma$ 
    and $C_{\text{R}} = C_{\text{L}} = 100 C_{\text{g}}$.
    (b): Same as (a) except for the finite Zeeman splitting by $h = 15T$
    and asymmetric capacitances
    $C_{\text{L}} = 0.7 C_{\text{R}} = 100 C_{\text{g}}$.
    In addition, the gate capacitance of state 2 is larger:
    $C_{\text{g},2} = 1.176 C_{\text{g}}$.
    (c): $d^2I/dV^2$ vs $V$ around the PT threshold $V_\text{PT}$ at 
    $V_\text{g} = -0.625 U C / C_\text{g}$
    ($C = C_\text{L} + C_\text{R} + C_\text{g}$),
    indicated  by the red line in (a).
    This demonstrates the scaling with $1/U^2$ and the agreement with
    the analytic expression Eq.~(\ref{eq:analyticpeak})  marked by (A).
    (d): The two independent Keldysh diagrams contributing to the pair-tunneling resonance.
    A sum over $\sigma$ is assumed and $\bar \sigma$ denotes the opposite spin-projection of 
    $\sigma$.
  }
\end{figure}

Fig.~\ref{fig:anderson} (a) and (b) show the differential conductance as a function of gate and bias voltage,
where in (b) the spin-degeneracy is lifted by an applied magnetic field.
SET processes give rise to the strong yellow lines which cross at zero
bias at the charge degeneracy points. 
In (b) additional gate-independent steps show up in the $N = 1$ Coulomb blockade region, 
indicating the onset of inelastic
cotunneling leading to occupation of the excited spin-state. Subsequent relaxation
of this state by SET gives rise to additional gate-dependent lines in the Coulomb blockade
region, as described e.g. in Refs.~\cite{Golovach04,Schleser05}.
Clearly separated from these other resonances, the PT resonance 
associated with an electron pair tunneling onto (off) the dot
appears as a gate-dependent step inside the SET region, indicated by a red solid (green dotted) arrow in (a).
Considering the resonance position and its voltage dependence only,
the PT could be mistaken for a weak SET process involving an excited
state of the $N=1$ charge state with energy $\Delta=U/2$ relative to the $N=1$
ground state.
However, the resonance shape provides one crucial clue to the identification as PT:
in Fig.~\ref{fig:anderson}(c) we show a bias trace along the red vertical line 
indicated in Fig~\ref{fig:anderson}(a), where it is seen that the PT resonance appears 
as a peak in $d^2I/dV^2$, rather than in $dI/dV$ as for SET. 
The reason is that, analogous to inelastic cotunneling, the number of electron states 
in the reservoirs available for a PT process is proportional to the bias voltage~\cite{Lambe68}.
Another clear way to separate PT and SET resonances is that the PT resonance exhibits no Zeeman splitting, 
see Fig.~\ref{fig:anderson}(b).
The physical reason is that the tunneling of a pair involves a
transition between states with the same spin ($N=0,2$ electron
singlets), expressed by the appearance of the sum $E_{\uparrow} +
E_{\downarrow}$ in Eq.~(\ref{eq:resonance_condition}).
Since PT relies on an empty or doubly occupied initial state, the resonances do not continue down into 
the linear transport regime, making PT an \emph{exclusively non-equilibrium effect}.
Additionally, in contrast to the magnetic field excitations, the PT resonances do not continue horizontally 
into the $N = 1$ Coulomb blockade region as inelastic cotunneling steps.
In cases which deviate from the simplest capacitive model, the PT resonance shows an additional 
distinct property. In Fig.~\ref{fig:anderson}(b) the capacitances associated with the left and right 
leads were chosen unequal and the doubly occupied state has a larger gate-coupling 
(this might happen if the \emph{many-body} wave-function of the doubly occupied state is localized
closer to the gate electrode, or is less screened by the source and drain electrodes~\cite{Osorio07b}). 
This causes the diamond to be both tilted and skewed, but the PT resonance can still be found by taking the 
average bias voltage positions (c.f.~(\ref{eq:resonance_condition})) of the $N = 0 \leftrightarrow N = 1$ and $N = 1 \leftrightarrow N = 2$ ground state 
transitions respectively.
\par
The precise amplitude and shape of the PT peak are also crucial for its correct identification.
Here an analytic expression for the peak is helpful, which can be obtained for 
$U \gg T, \Gamma, h$. 
For this we focus on the upper left SET region in Fig.~\ref{fig:anderson}(a--b) ($V > 0$) where the PT
involves the transition $|0\rangle \rightarrow |2\rangle$ with both electrons tunneling onto the dot 
from the left reservoir.
Deep in the SET region the SET rates are constant, while all second order rates except the pair-tunneling 
are approximately linear in bias voltage. Here the PT peak is given by
\begin{eqnarray}
\label{eq:pairpeak_1}
	\frac{d^2 I_\text{L}^{\text{PT}}}{d V^2} &=& 
	2 \left( \frac{ \Gamma_\text{R} }{ 2 \Gamma_\text{L} + 
	\Gamma_\text{R} } \right)^2 \frac{d^2}{d V^2} W_{2 0}^\text{LL}.
\end{eqnarray}
There are only two independent terms in the perturbation expansion which contribute to the {non-linear
voltage dependence of} the rate for tunneling with an electron pair from the left reservoir, 
$W_{2 0}^\text{LL} = \text{Re} ( X_{2 0}^\text{LL} + D_{2 0}^\text{LL} )$.
These terms correspond to the real-time diagrams~\cite{Koenig96prl} shown in Fig~\ref{fig:anderson}(d).
The corresponding integrals can be solved analytically~\cite{Leijnse08a}
without the need for ad hoc regularization. 
Eq.~(\ref{eq:pairpeak_1}) shows that the peak in the derivative of the differential conductance 
directly maps out the energy dependence of the pair-tunneling rate.
Neglecting all terms not contributing to the PT peak in the second derivative, 
assuming equal capacitances to the left and right reservoirs,
we are left with
\begin{eqnarray}
\label{eq:analyticpeak}
        \frac{d^2 I_\text{L}^{\text{PT}}}{dV^2}
	&=&
         \frac{4 \Gamma_\text{L} \Gamma_\text{R} }{\pi T U^2}  
	\frac{ \Gamma_\text{L} \Gamma_\text{R} }{ \left( 2 \Gamma_\text{L} + \Gamma_\text{R} \right)^2 }
	F \left( \frac{E_{\uparrow} + E_{\downarrow} + U - V}{T}\right) \nonumber \\ 
	&\times&
	\sum_{s = \pm} \frac{1}{1 + s h / U} 
	\left( 1 + \frac{1}{1 + s h / U} \right).
\end{eqnarray}
The resonance condition~(\ref{eq:resonance_condition}) is seen in the argument of the function
$F(x) = d^2  \left( x b(x) \right) / dx^2 
=  \left( \frac{x}{4} \coth \frac{x}{2} - \frac{1}{2} \right) / \sinh^2 \left( \frac{x}{2} \right) $, 
with $b(x) = 1/\left( e^{x} - 1 \right)$ being the Bose function and $F(0) = 1/6$.
\par
{It is of interest to compare the PT with the inelastic COT, which is theoretically and experimentally well-studied.
In general the PT and COT peaks are of comparable magnitude and their dependence on $T$ and $U$ are the same,
 implying that the PT and COT should be simultaneously experimentally accessible.
A first difference is that the PT peak is completely insensitive to relaxation processes,
since it involves a transition between ground states in different charge sectors.
In contrast, the COT resonance consists of two parts.
The part which  remains even in the limit where the voltage dependence of the occupations is 
negligible (relaxation faster  than the electron cotunneling and / or asymmetric junction
$\Gamma_{L,R} \gg \Gamma_\text{R,L}$), is given by:
}
\begin{eqnarray}
\label{eq:analyticCOTpeak}
        \frac{d^2 I^{\text{COT}}_\text{L}}{dV^2}
	&=&
        \frac{8 \Gamma_\text{L} \Gamma_\text{R} }{\pi T U^2}  
	F \left( \frac{h - V}{T}\right). 
\end{eqnarray}
This expression is valid  at the particle hole symmetric point (center of the $N=1$ 
Coulomb blockade regime) in the limit $U \gg h > T > \Gamma$ and $V>0$.
For slow relaxation and $\Gamma_\text{L} \approx \Gamma_\text{R}$ the COT peak-height is additionally increased and also an 
asymmetry between $V > h$ and $V < h$ appears due to the voltage dependence of the 
occupations~\cite{Paaske04Kondo}.
For very small fields and $\Gamma_\text{L} = \Gamma_\text{R}$, 
the COT peak height is seen to be given by $9/2$ times that of the PT peak.
A second difference is that the PT peak is strongly suppressed for very asymmetric tunnel rates,
proportional to the square of the smaller coupling and,
characteristically, the forward and  reverse bias PT peak-heights differ by a factor 4,
 in contrast to the COT peaks which remain symmetric.
Finally, although a magnetic field does not cause a Zeeman splitting of the PT 
peak, Eq.~(\ref{eq:analyticpeak}),
it does increase the peak height, in contrast to the COT peak, Eq.~(\ref{eq:analyticCOTpeak}), 
where the situation is reversed:
the peak height is independent of the field, while the position 
($V = h$, independent of $V_\text{g}$) is shifted linearly.
\par
Quantum dots with comparable level spacing and charging energy
 are especially well suited for observation of PT since the background SET current is 
featureless.
In the case of a dense excitation spectrum, e.g. due to coupling to a localized vibrational 
mode, additional PT resonances complicate a high-resolution spectroscopic analysis. 
{Extending the calculations of the non-equilibrium Anderson-Holstein model} 
in Ref.~\cite{Leijnse08a} 
{to finite $U$ we find that}
PT resonances give a signature similar to SET associated with a {very} weakly coupled 
second vibrational mode.  
Thus, the results presented here are of general importance for the analysis of models
dealing with the various complexities of realistic quantum dots. 
In the non-interacting limit, $U = 0$, the PT is completely suppressed.
In general, PT involves interference between the two contributions (diagrams) 
shown in Fig.~\ref{fig:anderson}(d).
This interference may be constructive, as in the Anderson model, or destructive 
in other models. 
Interestingly the relative sign of the interfering contributions
is sensitive to the spin of the involved quantum dot states.
For a generalized quantum dot model where the doubly occupied ground state is a triplet,
we find that the PT {is suppressed} due to quantum interference and can be {induced} 
by a magnetic field.
Finally, 
the PT also shows up in more complex transport quantities such as the current noise, which is sensitive to the effective charge 
transferred in the tunnel process (which is $\pm 2e$ here). 
Extending the real-time approach to the calculation of shot-noise~\cite{Thielmann05, Aghassi08}, including 
non-Markovian effects, we find that the PT resonance
is associated with an increased Fano factor.
The above examples serve to illustrate that pair-tunneling effects are of general importance for transport through nanosystems.
\par
{\em Discussion and conclusion.}
In this letter we have theoretically predicted a signature of coherent tunneling of electron pairs 
in the single-electron transport regime,
for quantum dots with positive charging energies exceeding the broadening by thermal and quantum fluctuations of charge.
Current low-temperature measurements can access such fine details in the first three
derivatives of the current with respect to voltage without dropping below the noise level~\cite{Osorio07b}.
Also, many molecular and carbon nanotube devices couple strongly to the electrodes, making 
first and second order transport features comparable.
We mention the possibility of measuring the charging energy by identifying the PT resonance 
and reading U/2 off in the stability diagram,
thus limiting the voltages one needs to apply.
Alternatively, the pair tunneling resonance provides a consistency check on SET level assignments.
On a general level, the results indicate the importance of complete perturbative treatment of non-equilibrium problems~\cite{Leijnse08a}.
\par
We acknowledge many stimulating discussions with
H. Schoeller,
J. K\"onig
and F. Reckermann
and the financial support from
DFG SPP-1243,
the NanoSci-ERA, the Helmholtz Foundation and
the FZ-J\"ulich (IFMIT).

\bibliographystyle{apsrev}

\begin{thebibliography}{15}
\expandafter\ifx\csname natexlab\endcsname\relax\def\natexlab#1{#1}\fi
\expandafter\ifx\csname bibnamefont\endcsname\relax
  \def\bibnamefont#1{#1}\fi
\expandafter\ifx\csname bibfnamefont\endcsname\relax
  \def\bibfnamefont#1{#1}\fi
\expandafter\ifx\csname citenamefont\endcsname\relax
  \def\citenamefont#1{#1}\fi
\expandafter\ifx\csname url\endcsname\relax
  \def\url#1{\texttt{#1}}\fi
\expandafter\ifx\csname urlprefix\endcsname\relax\def\urlprefix{URL }\fi
\providecommand{\bibinfo}[2]{#2}
\providecommand{\eprint}[2][]{\url{#2}}

\bibitem[{\citenamefont{Kouwenhoven et~al.}(1997)\citenamefont{Kouwenhoven,
  Marcus, McEuen, Tarucha, Westervelt, and Wingreen}}]{Kouwenhoven97rev}
\bibinfo{author}{\bibfnamefont{L.}~\bibnamefont{Kouwenhoven}},
  \bibinfo{author}{\bibfnamefont{C.}~\bibnamefont{Marcus}},
  \bibinfo{author}{\bibfnamefont{P.}~\bibnamefont{McEuen}},
  \bibinfo{author}{\bibfnamefont{S.}~\bibnamefont{Tarucha}},
  \bibinfo{author}{\bibfnamefont{R.}~\bibnamefont{Westervelt}},
  \bibnamefont{and} \bibinfo{author}{\bibfnamefont{N.}~\bibnamefont{Wingreen}},
  in \emph{\bibinfo{booktitle}{Mesoscopic electron transport}}, edited by
  \bibinfo{editor}{\bibfnamefont{K.}~\bibnamefont{Sohn}},
  \bibinfo{editor}{\bibfnamefont{L.}~\bibnamefont{Kouwenhoven}},
  \bibnamefont{and} \bibinfo{editor}{\bibfnamefont{G.}~\bibnamefont{Sch\"on}}
  (\bibinfo{publisher}{Kluwer}, \bibinfo{year}{1997}), chap.
  \bibinfo{chapter}{Electron Transport in Quantum Dots}.

\bibitem[{\citenamefont{Lambe and Jaklevic}(1968)}]{Lambe68}
\bibinfo{author}{\bibfnamefont{J.}~\bibnamefont{Lambe}} \bibnamefont{and}
  \bibinfo{author}{\bibfnamefont{R.~C.} \bibnamefont{Jaklevic}},
  \bibinfo{journal}{Phys.\ Rev.} \textbf{\bibinfo{volume}{165}},
  \bibinfo{pages}{821} (\bibinfo{year}{1968}).

\bibitem[{\citenamefont{DeFranceschi et~al.}(2001)\citenamefont{DeFranceschi,
  Sasaki, Elzerman, van~der Wiel, Tarucha, and Kouwenhoven}}]{DeFranceschi01}
\bibinfo{author}{\bibfnamefont{S.}~\bibnamefont{DeFranceschi}},
  \bibinfo{author}{\bibfnamefont{S.}~\bibnamefont{Sasaki}},
  \bibinfo{author}{\bibfnamefont{J.~M.} \bibnamefont{Elzerman}},
  \bibinfo{author}{\bibfnamefont{W.~G.} \bibnamefont{van~der Wiel}},
  \bibinfo{author}{\bibfnamefont{S.}~\bibnamefont{Tarucha}}, \bibnamefont{and}
  \bibinfo{author}{\bibfnamefont{L.~P.} \bibnamefont{Kouwenhoven}},
  \bibinfo{journal}{Phys.\ Rev.\ Lett.} \textbf{\bibinfo{volume}{86}},
  \bibinfo{pages}{878} (\bibinfo{year}{2001}).

\bibitem[{\citenamefont{Sela et~al.}(2008)\citenamefont{Sela, Sim, Oreg, Raikh,
  and von Oppen}}]{Sela08}
\bibinfo{author}{\bibfnamefont{E.}~\bibnamefont{Sela}},
  \bibinfo{author}{\bibfnamefont{H.-S.} \bibnamefont{Sim}},
  \bibinfo{author}{\bibfnamefont{Y.}~\bibnamefont{Oreg}},
  \bibinfo{author}{\bibfnamefont{M.~E.} \bibnamefont{Raikh}}, \bibnamefont{and}
  \bibinfo{author}{\bibfnamefont{F.}~\bibnamefont{von Oppen}},
  \bibinfo{journal}{Phys.\ Rev.\ Lett.} \textbf{\bibinfo{volume}{100}},
  \bibinfo{pages}{056809} (\bibinfo{year}{2008}).

\bibitem[{\citenamefont{Koch et~al.}(2006)\citenamefont{Koch, Raikh, and von
  Oppen}}]{Koch05c}
\bibinfo{author}{\bibfnamefont{J.}~\bibnamefont{Koch}},
  \bibinfo{author}{\bibfnamefont{M.~E.} \bibnamefont{Raikh}}, \bibnamefont{and}
  \bibinfo{author}{\bibfnamefont{F.}~\bibnamefont{von Oppen}},
  \bibinfo{journal}{Phys.\ Rev.\ Lett.} \textbf{\bibinfo{volume}{96}},
  \bibinfo{pages}{056803} (\bibinfo{year}{2006}).

\bibitem[{\citenamefont{Hekking et~al.}(1993)\citenamefont{Hekking, Glazman,
  Matveev, and Shekhter}}]{Hekking93}
\bibinfo{author}{\bibfnamefont{F.~W.~J.} \bibnamefont{Hekking}},
  \bibinfo{author}{\bibfnamefont{L.~I.} \bibnamefont{Glazman}},
  \bibinfo{author}{\bibfnamefont{K.~A.} \bibnamefont{Matveev}},
  \bibnamefont{and} \bibinfo{author}{\bibfnamefont{R.~I.}
  \bibnamefont{Shekhter}}, \bibinfo{journal}{Phys.\ Rev.\ Lett.}
  \textbf{\bibinfo{volume}{70}}, \bibinfo{pages}{4138} (\bibinfo{year}{1993}).

\bibitem[{\citenamefont{Cornaglia and Grempel}(2005)}]{Cornaglia05b}
\bibinfo{author}{\bibfnamefont{P.~S.} \bibnamefont{Cornaglia}}
  \bibnamefont{and} \bibinfo{author}{\bibfnamefont{D.~R.}
  \bibnamefont{Grempel}}, \bibinfo{journal}{Phys.\ Rev.\ B}
  \textbf{\bibinfo{volume}{71}}, \bibinfo{pages}{245326}
  (\bibinfo{year}{2005}).

\bibitem[{\citenamefont{K\"onig et~al.}(1996)\citenamefont{K\"onig, Schoeller,
  and Sch\"on}}]{Koenig96prl}
\bibinfo{author}{\bibfnamefont{J.}~\bibnamefont{K\"onig}},
  \bibinfo{author}{\bibfnamefont{H.}~\bibnamefont{Schoeller}},
  \bibnamefont{and} \bibinfo{author}{\bibfnamefont{G.}~\bibnamefont{Sch\"on}},
  \bibinfo{journal}{Phys.\ Rev.\ Lett.} \textbf{\bibinfo{volume}{76}},
  \bibinfo{pages}{1715} (\bibinfo{year}{1996}).

\bibitem[{\citenamefont{Leijnse and Wegewijs}(2008)}]{Leijnse08a}
\bibinfo{author}{\bibfnamefont{M.}~\bibnamefont{Leijnse}} \bibnamefont{and}
  \bibinfo{author}{\bibfnamefont{M.~R.} \bibnamefont{Wegewijs}},
  \bibinfo{journal}{Phys.\ Rev.\ B} \textbf{\bibinfo{volume}{78}},
  \bibinfo{pages}{235424} (\bibinfo{year}{2008}), \bibinfo{note}{in Fig. 1(a)
  of this paper the PT resonance, although present, is not resolved with the
  color scale used.}

\bibitem[{\citenamefont{Golovach and Loss}(2004)}]{Golovach04}
\bibinfo{author}{\bibfnamefont{V.~N.} \bibnamefont{Golovach}} \bibnamefont{and}
  \bibinfo{author}{\bibfnamefont{D.}~\bibnamefont{Loss}},
  \bibinfo{journal}{Phys.\ Rev.\ B} \textbf{\bibinfo{volume}{69}},
  \bibinfo{pages}{245327} (\bibinfo{year}{2004}).

\bibitem[{\citenamefont{Schleser et~al.}(2005)\citenamefont{Schleser, Ihn, Ruh,
  Ensslin, Tews, Pfannkuche, Driscoll, and Gossard}}]{Schleser05}
\bibinfo{author}{\bibfnamefont{R.}~\bibnamefont{Schleser}},
  \bibinfo{author}{\bibfnamefont{T.}~\bibnamefont{Ihn}},
  \bibinfo{author}{\bibfnamefont{E.}~\bibnamefont{Ruh}},
  \bibinfo{author}{\bibfnamefont{K.}~\bibnamefont{Ensslin}},
  \bibinfo{author}{\bibfnamefont{M.}~\bibnamefont{Tews}},
  \bibinfo{author}{\bibfnamefont{D.}~\bibnamefont{Pfannkuche}},
  \bibinfo{author}{\bibfnamefont{D.~C.} \bibnamefont{Driscoll}},
  \bibnamefont{and} \bibinfo{author}{\bibfnamefont{A.~C.}
  \bibnamefont{Gossard}}, \bibinfo{journal}{Phys.\ Rev.\ Lett.}
  \textbf{\bibinfo{volume}{94}}, \bibinfo{pages}{206805}
  (\bibinfo{year}{2005}).

\bibitem[{\citenamefont{Osorio et~al.}(2007)\citenamefont{Osorio, O'Neill,
  Wegewijs, Stuhr-Hansen, Paaske, Bj{\o}rnholm, and van~der Zant}}]{Osorio07b}
\bibinfo{author}{\bibfnamefont{E.~A.} \bibnamefont{Osorio}},
  \bibinfo{author}{\bibfnamefont{K.}~\bibnamefont{O'Neill}},
  \bibinfo{author}{\bibfnamefont{M.~R.} \bibnamefont{Wegewijs}},
  \bibinfo{author}{\bibfnamefont{N.}~\bibnamefont{Stuhr-Hansen}},
  \bibinfo{author}{\bibfnamefont{J.}~\bibnamefont{Paaske}},
  \bibinfo{author}{\bibfnamefont{T.}~\bibnamefont{Bj{\o}rnholm}},
  \bibnamefont{and} \bibinfo{author}{\bibfnamefont{H.~S.} \bibnamefont{van~der
  Zant}}, \bibinfo{journal}{Nanolett.} \textbf{\bibinfo{volume}{7}},
  \bibinfo{pages}{3336} (\bibinfo{year}{2007}).

\bibitem[{\citenamefont{Paaske et~al.}(2004)\citenamefont{Paaske, Rosch, and
  W\"olfle}}]{Paaske04Kondo}
\bibinfo{author}{\bibfnamefont{J.}~\bibnamefont{Paaske}},
  \bibinfo{author}{\bibfnamefont{A.}~\bibnamefont{Rosch}}, \bibnamefont{and}
  \bibinfo{author}{\bibfnamefont{P.}~\bibnamefont{W\"olfle}},
  \bibinfo{journal}{Phys.\ Rev.\ B} \textbf{\bibinfo{volume}{69}},
  \bibinfo{pages}{155330} (\bibinfo{year}{2004}).

\bibitem[{\citenamefont{Thielmann et~al.}(2005)\citenamefont{Thielmann,
  Hettler, K\"onig, and Sch\"on}}]{Thielmann05}
\bibinfo{author}{\bibfnamefont{A.}~\bibnamefont{Thielmann}},
  \bibinfo{author}{\bibfnamefont{M.~H.} \bibnamefont{Hettler}},
  \bibinfo{author}{\bibfnamefont{J.}~\bibnamefont{K\"onig}}, \bibnamefont{and}
  \bibinfo{author}{\bibfnamefont{G.}~\bibnamefont{Sch\"on}},
  \bibinfo{journal}{Phys.\ Rev.\ Lett.} \textbf{\bibinfo{volume}{95}},
  \bibinfo{pages}{146806} (\bibinfo{year}{2005}).

\bibitem[{\citenamefont{Aghassi et~al.}(2008)\citenamefont{Aghassi, Hettler,
  and Sch\"on}}]{Aghassi08}
\bibinfo{author}{\bibfnamefont{J.}~\bibnamefont{Aghassi}},
  \bibinfo{author}{\bibfnamefont{M.}~\bibnamefont{Hettler}}, \bibnamefont{and}
  \bibinfo{author}{\bibfnamefont{G.}~\bibnamefont{Sch\"on}},
  \bibinfo{journal}{APL} \textbf{\bibinfo{volume}{92}}, \bibinfo{pages}{202101}
  (\bibinfo{year}{2008}).

\end{thebibliography}

\end{document}